\documentclass[]{article}
\usepackage {graphicx}
\usepackage{amssymb}
\usepackage[english]{babel}
\newcommand{\bea}{\begin{eqnarray}}
\newcommand{\eea}{\end{eqnarray}}
\newcommand{\be}{\begin{equation}}
\newcommand{\ee}{\end{equation}}
\newcommand{\beaa}{\begin{eqnarray*}}
\newcommand{\eeaa}{\end{eqnarray*}}
\newcommand{\eps}{\varepsilon}

\begin{document}
\sloppy

\title{Influence of external factors on dielectric permittivity of
Rochelle salt: humidity, annealing, stresses, electric field}
\author{A.G.Slivka$^*$, V.M.Kedyulich$^*$,
R.R.Levitskii$^{**}$, \\
A.P.Moina$^{**}$,
 M.O.Romanyuk$^{***}$, A.M. Guivan$^*$\\
$^*$ Uzhgorod National University, 54 Voloshin Street, Uzhgorod,
Ukraine\\
$^{**}$ Institute for Condensed Matter Physics, 1 Svientsitskii
Street, Lviv,  Ukraine\\
$^{***}$Ivan Franko Lviv National University, 8 Kyryla and
Mefodiya Street, Lviv,
 Ukraine}
\date{}
 \maketitle

\begin{abstract}
The present work contains results of experimental
 investigation of external factors, such as dessicating/wetting,
 thermal annealing, uniaxial and hydrostatic pressures on
 dielectric permittivity of Rochelle salt crystals. The obtained
 results are compared with available literature data. A conslusion
 is made that the dispersion of experimental data can be
 attributed to internal polar point defects in crystals and to
 influence of storage conditions. The obtained results are analyzed  within the phenomenological Landau
 approach.

\end{abstract}

\section{Introduction}
An important information about the transition mechanism in ferroelectric
crystals can be given by exploring their behavior under influence of various
external factors, such as high pressure or electric field.
For hydrogen bonded crystals the external pressures are the only way to
continuously vary geometric parameters of bonds, break their equivalence,
etc, which permits to investigate role of hydrogen bonds and their parameters and symmetry
in mechanisms of the phase transition and dielectric response of the crystals.
Many of ferroelectrics are piezoelectric in the paraelectric phase; application of
shear stresses and the conjugate electric fields provides a possibility to explore
the role of piezoelectric interactions in the phase transitions and in formation
of the physical characteristics of the crystals.

The above mentioned possibilities were fully used for investigation of the
KH$_2$PO$_4$ family crystals. Theoretical description of pressure and field
effects in these crystals are usually performed within the proton ordering model
(see e.g. \cite{our!,our!!,duda,Slivka,our13,ferro} and references therein) and a
quantitative agreement with experiment is obtained. It was shown, in
particular, that pressures of different symmetries can produce
qualitatively different changes in the phase transition: lower its
temperature down (hydrostatic), raise it up and smear out the
transition (as shear stress $\sigma_6$), induce a new phase of
monoclinic symmetry (as $\sigma_1-\sigma_2$).

In contrast to the KH$_2$PO$_4$ family crystals, the data for
external factors influence on Rochelle salt crystals are less
extensive. In literature, only a few papers are available on
hydrostatic pressure  \cite{slivka8,slivka11,slivka12} and
electric field \cite{slivka10,slivka6,slivka13} effects on the
dielectric permittivity of the crystals. Uniaxial stresses effects
on the phase transitions in Rochelle salt were explored in in
\cite{imai} from the measurements of thermoelastic effect.
Theoretically influence of the shear stress $\sigma_4$ was studied
in \cite{sigma4} within the modified Mitsui model.

Usually, peculiarities of the physical characteristics of ferroelectric crystals
in the vicinity of the phase transitions (especially of the second order ones) are affected by crystal
defects and internal bias electric fields and mechanical stresses,
which act as the external ones. The role of such factors, as
crystal defectness and the processes in the sample prehistory, that may affect
the physical properties of the crystals via relaxation of the defects: thermal annealing, previous
influence of electromagnetic fields and mechanical stresses, must
be explored. High pressure and electric
field studies allow to explore the intrinsic field and pressure
dependences of the crystal properties, reveal the internal bias fields and stresses, and study
the residual effects of the internal defects.

For Rochelle salt, whose chemical instability (loss of crystallization
water at slightly elevated temperatures) and water absorbency are well
known, other factors such as air humidity, storage conditions, etc are
important and should be monitored during measurements. For instance,
a significant dispersion of experimental data for the dielectric
permittivity of Rochelle salt (see the systematization in \cite{slivka1}),
exceeding the measurement error, takes place. Apparently, the
dispersion is due to the different internal states of the samples,
not controlled during measurements.

In the present work the results of experimental studies of the
mentioned above external factors (pressure, electric field, humidity, thermal annealing)
influence on dielectric permittivity of Rochelle salt crystals in
the vicinity of the structural phase transitions are reported.

\section{Experimental setup}
Dielectric permittivity of the crystals $\varepsilon_{11}$ was
determined by measuring the samples capacity with the help of an
a.c. bridge at fixed frequency of 1 kHz. Measurement error was
$0.2\div0.4$\%.

Samples were prepared in a form of parallellepipeds, with faces
perpendicular to crystallographic axes of an orthorhombic
(paraelectric) unit cell. Silver paste and copper wires,
$0.08\div 0.12$~mm diameter, were used as electric contacts.
After partial drying of the paste, the contacts were covered by an alcohol solution
of a glue with addition of silver paste. This method provided a
necessary mechanical stability of the contacts and allowed a free deformation of the crystals.

A uniaxial pressure was created by a spring dynamometer and
transmitted to samples via a punch with floating heads, thus securing a uniform pressure even
at possibly non-parallel faces of the sample. The pressure was fixed with an accuracy of $\pm5$\%. The samples were placed in a thermostate, allowing smooth adjustment
of temperature. Temperature was measured by a copper-constantan
thermocouple with an accuracy  $\pm 0.1$ К.  Samples with the thermocouple were covered with silicone oil, in
order to enhance heat transmission and prevent direct contact with air.

\section{Model approach}
Theoretical description of the physical properties of Rochelle
salt is usually performed within a two-sublattice Ising model with
asymmetry double well potential (Mitsui model). Below we present
the expression for the dielectric permittivity of Rochelle salt
obtained within the modified Mitsui model with taking into account
the piezoelectric coupling \cite{slivka1}
with the Hamiltonian
 \bea &&\hat H = \frac{N}{2} {v}
c_{44}^{E0}\varepsilon_4^2 - N{v}e_{14}^0\varepsilon_4  E_1 -
\frac{N}{2} {v}\chi_{11}^0E_1^2- \frac12 \sum\limits_{qq'}\sum\limits_{ff'=1}^2
R_{qq'}(ff') \frac{\sigma_{qf}}{2}\frac{\sigma_{q'f'}}{2}
\nonumber\\
&& \label{2.1}
\quad  -\Delta \sum\limits_q \left(\frac{\sigma_{q_1}}{2}
-\frac{\sigma_{q_2}}{2} \right)- (\mu_1E_1-2\psi_4\varepsilon_4)
\sum\limits_q\sum\limits_{f=1}^2 \frac{\sigma_{qf}}{2}.
 \nonumber
\eea
Three first terms in (\ref{2.1}) correspond to a `seed'' energy of the
crystal lattice which  forms the asymmetric double-well potential
for the pseudospins. $R_{qq'}(11) = R_{qq'}(22) = J_{qq'}$ and
$R_{qq'}(12) = R_{qq'}(21) = K_{qq'}$ are constants of interaction
between pseudospins belonging to the same and to different
sublattices, respectively. The parameter $\Delta$ describes the
asymmetry of the double well potential; $\mu_1$ is the effective
dipole moment. The last term is the internal field created by the
piezoelectric coupling with the shear strain $\eps_4$.

Within a mean field approximation the static dielectric permittivity of a free
crystal is obtained in the form \cite{slivka1}
 \be \label{2.43}
\chi_{11}^{\sigma} =  \chi_{11}^{\sigma 0} +
  \frac{\beta(\mu_1')^2}{2v}
F_2(0),
\ee where $\xi$, $\sigma$ are the parameters of ferroelectric and antiferroelectric ordering.
The following notations are used
 \beaa
& F_2(0)=\frac{\displaystyle
\varphi_3}
{\displaystyle\varphi_2- \Lambda \varphi_3},&\\
&\varphi_2=1-\frac{\beta J}{2}\lambda_1- \beta^2\frac{K^2-
J^2}{16}(\lambda_1^2-\lambda_2^2),\quad
\varphi_3=\lambda_1+\beta\frac{K-J}{4}(\lambda_1^2-\lambda_2^2),\quad
\Lambda=\frac{\displaystyle2\beta\psi_4^2}{\displaystyle
vc_{44}^{E0}},&\\
& \lambda_1=1-\xi^2-\sigma^2, \qquad  \lambda_2=2\xi\sigma,&
\\& d_{14}^0 = \frac{\displaystyle e_{14}^0}{\displaystyle c_{44}^{E0}},\quad \chi_{11}^{\sigma
0} = \chi_{11}^{\varepsilon 0} + e_{14}^0d_{14}^0, \quad {\mu_1'}
= {\mu_1} - {2} \psi_4 d_{14}^0 .
&\eeaa
Values of the model parameters providing the best fit to the permittivity are given in Table~\ref{table}.

\begin{table}[tbh]
\caption{Model parameters for Rochelle salt
\protect\cite{slivka1}.}
\begin{center}
   \begin{tabular}{ccccccc}
     \hline
 $J/k_B$ & $K/k_B$ & $
     \Delta/k_B$ & $\psi_4/k_B$ & $c_{44}^{E0}$ & $d_{14}^0$ & $\chi^{\sigma0}_{11}$
    \\        \multicolumn{4}{c}{K} & dyn/cm$^2$ & esu/dyn &
     \\
     \hline  797.36 & 1468.83 &   737.33        &    -760   &
     $12.8\cdot 10^{10}$ & $1.9\cdot10^{-8}$ &0.363  \\
\hline \end{tabular}
     \end{center}
$v=0.5219 [1+ 0.00013(T-190)]\cdot 10^{-21}~{\rm cm}^{3}$
\label{table}
     \end{table}

%
%
%
%
%
%
%

\section{Influence of sample prehistory on dielectric permittivity of Rochelle salt}

\subsection{Humidity}

In \cite{slivka4} it was found that crystals of Rochelle salt at  25$^\circ$С
and relative humidity below 40\% lose the crystallization water, whereas at relative
humidity above 85\% they absorb water from air. Experimentally,
significant changes of the piezoelectric properties of Rochelle
salt were observed, when samples are kept in air with high
concentration of ethanol vapor \cite{slivka3}.

\begin{figure}[hbtp]
\centerline{\includegraphics[width=2.7in]{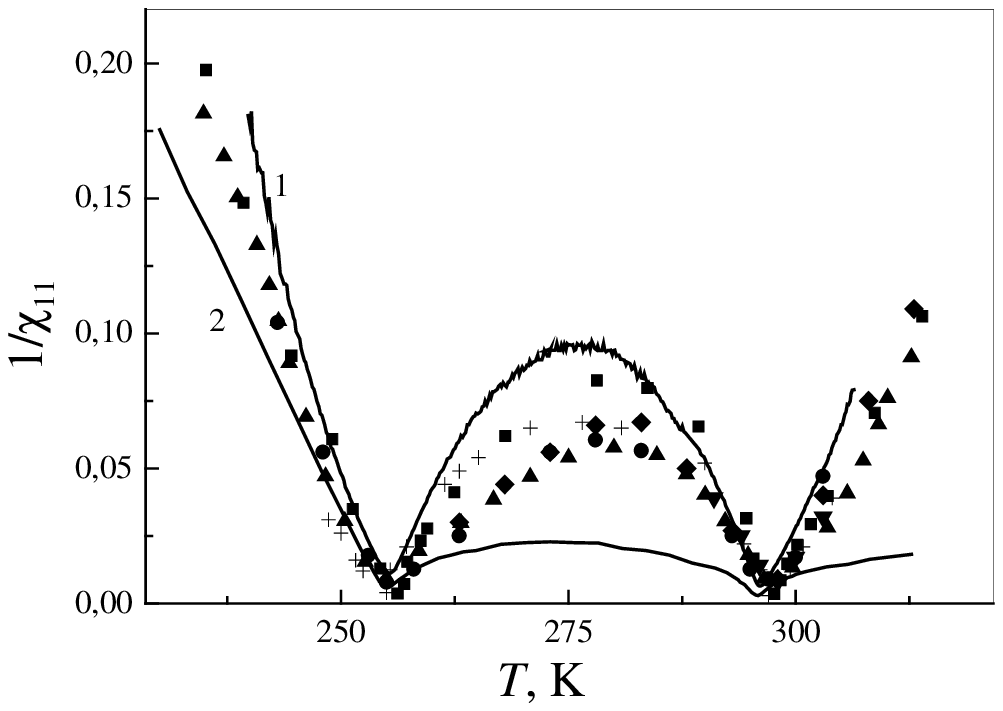}\hspace{2em}
\includegraphics[width=2.7in]{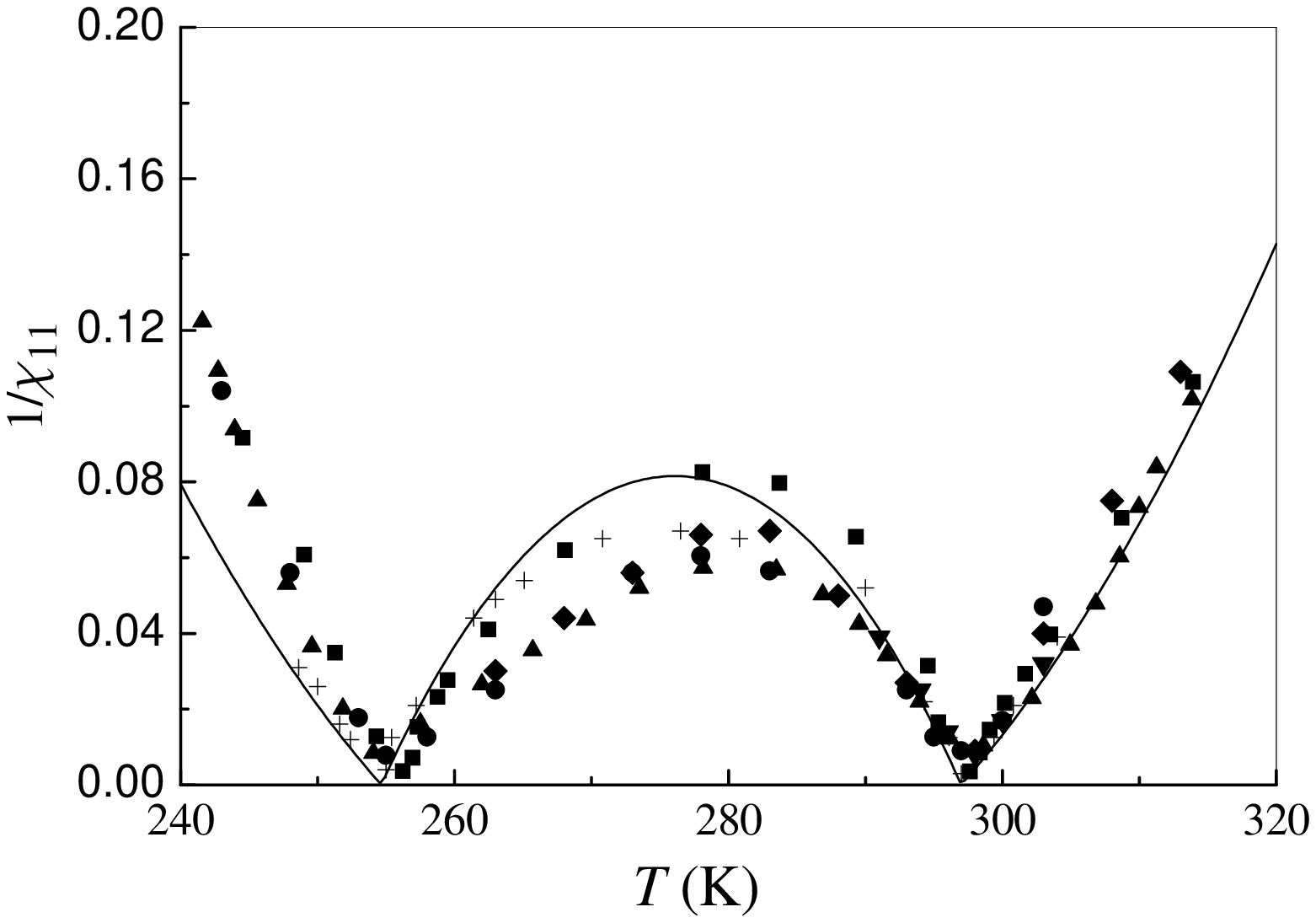}}
\caption{\small Temperature dependences of inverse susceptibility of Rochelle salt.
Left: solid lines -- experimental data, this work, obtained for 1
-- dessicated sample; 2 -- wet sample. Right: solid line -- theoretical
curve, calculated with  (\ref{2.43}). $\blacksquare$ -- \cite{15};
$\blacktriangle$ -- \cite{29}; $\blacklozenge$ -- \cite{24};
$\bullet$ -- \cite{27}; $\blacktriangledown$ -- \cite{28}; $+$ --
\cite{30}.} \label{slivka3}
\end{figure}

Experimental data for the susceptibility of Rochelle salt of different sources
(see fig.~\ref{slivka3}) show an essential dispersion, even in the paraelectric
phases, which cannot be accounted for by the changes in the
measurements regimes. Of interest was thus to explore the
temperature dependences of Rochelle salt crystals with different
water content, in order to verify whether this dispersion can be attributed, at least partially,
to it.

The obtained results are presented in fig.~\ref{slivka3} (left). Apart from the literature
data, we show here the temperature dependences of the inverse dielectric susceptibility
$\chi_{11}^{-1}$ (solid curves 1 and 2), obtained in this work for the same sample
with different water content. The curve~1 is obtained for a sample, kept for
a long time (2--3 days) at room temperature in a closed volume, filled with a dessicator (silicagel). The curve~2
corresponds to the same sample, kept for 10 hours in air with
relative humidity $\sim $90{\%}. As one can see,  keeping the
sample in a wet air decreases the dielectric susceptibility in the entire
studied temperature range. The changes are particularly prominent in the middle of the
ferroelectric phase $T\sim 275$~K and in the high-temperature paraelectric phase.

Comparison of the obtained results with literature data shows that the dispersion in the values of the susceptibility
indeed can be caused by a different water content in the samples
used in different experiments. It should be also noted that for a
wet sample (curve 2, fig.~\ref{slivka3}), a linear temperature dependence of inverse susceptibility
 $\chi^{-1}(T)$ with the Curie-Weiss constant $C_{W} =
1.95\cdot 10^3$ K. For a dessicated sample, the dependence $\chi^{-1}(T)$
is non-linear in both paraelectric phases.

Comparison of literature experimental data with the theoretical ones, obtained in \cite{slivka1}
from the formula (\ref{2.43}), is given in fig.~\ref{slivka3} (right). Theoretical absolute values of the permittivity are
adjusted by the choice of the value of the effective dipole moment $\mu_1$. In \cite{slivka1}
$\mu_1$  was chosen such as the best agreement with the data of  \cite{28} as well as of
the dynamic microwave permittivity is obtained. On this, we, however, failed to get an adequate
agreement with experiment for susceptibility in the low-temperature paraelectric phase \cite{slivka1}.

\subsection{Influence of thermal annealing}

\begin{figure}[htb]
\centerline{\includegraphics[angle=270,width=3.in]{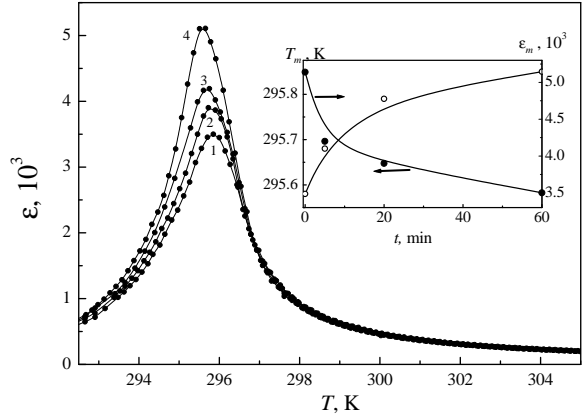}}
\caption{Temperature dependences of the dielectric permittivity of
Rochelle salt near the upper transition point at different times
of annealing  in the paraelectric phase at 308 K (min): 1 -- 0, 2
-- 5, 3 -- 20, 4 -- 60. Inset: dependences of the maximal value of
permittivity and maximum temperature on annealing time.}
\label{slivka4}
\end{figure}

Fit.~\ref{slivka4} illustrates the temperature dependences of dielectric permittivity
of Rochelle salt near the upper Curie point for samples annealed
at 308 K. On increasing the annealing time, the value
of the dielectric permittivity at the transition point increases, and the maximum temperature decreases.
Such changes are apparently caused by  internal
electrical bias fields, which magnitude is decreased with  increasing annealing time.

The internal bias fields are created by polar defects
\cite{Lines}, which at long-term keeping samples in the
ferroelectric phase participate in screening of spontaneous polarization and reflect the
corresponding domain structure. Action of the internal bias field is analogous to the action of external field, that is,
the temperature of the upper maximum of permittivity is increased, and the maximum magnitude is
decreased. In the next section we shall estimate the magnitudes of
the internal bias fields in the non-annealed and annealed samples.

%

\section{Influence of external electric field}

\begin{figure}[tbph]
\centerline{\includegraphics[width=2.in,angle=270]{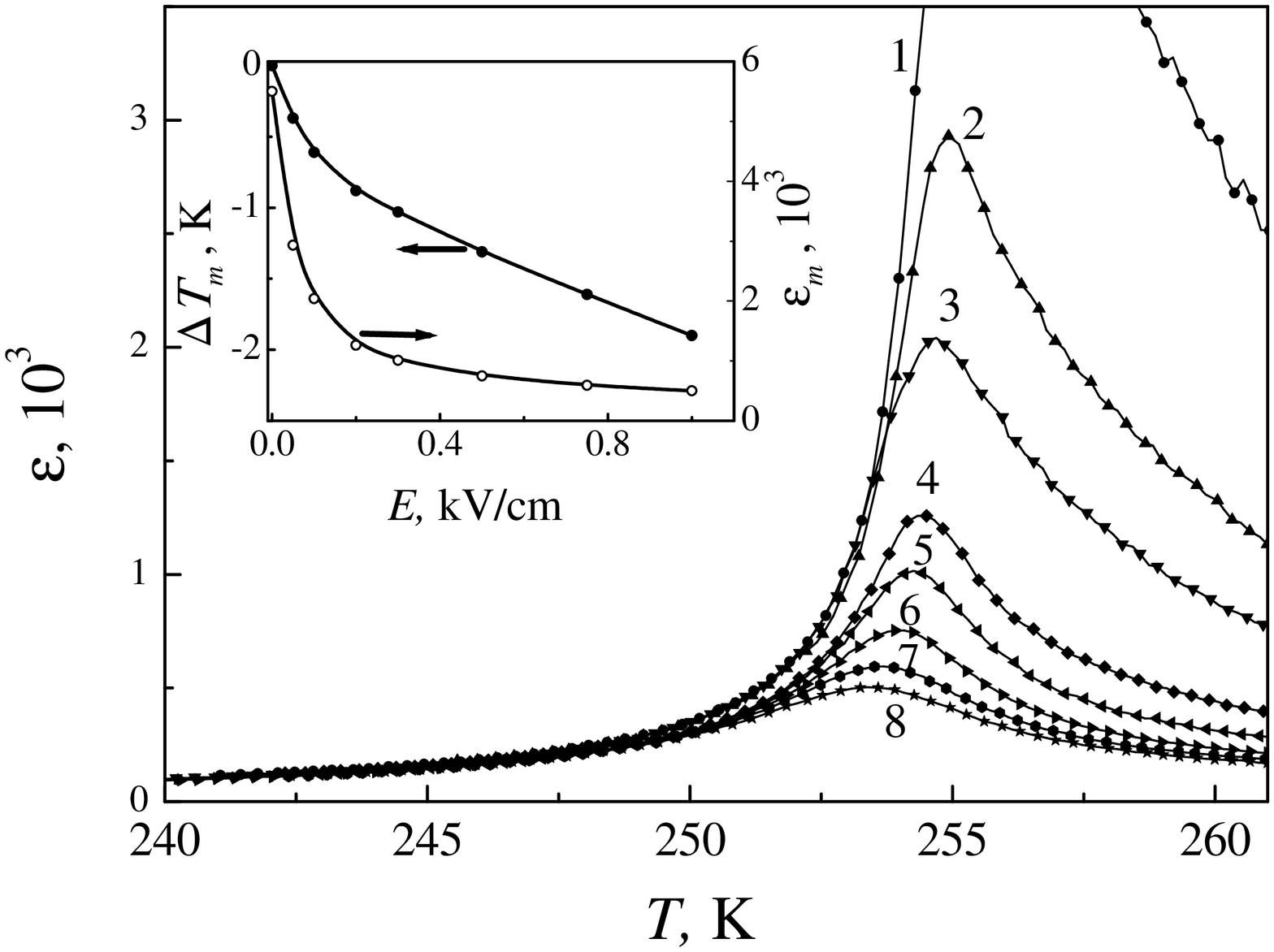}\hspace{2em}
\includegraphics[width=2.in,angle=270]{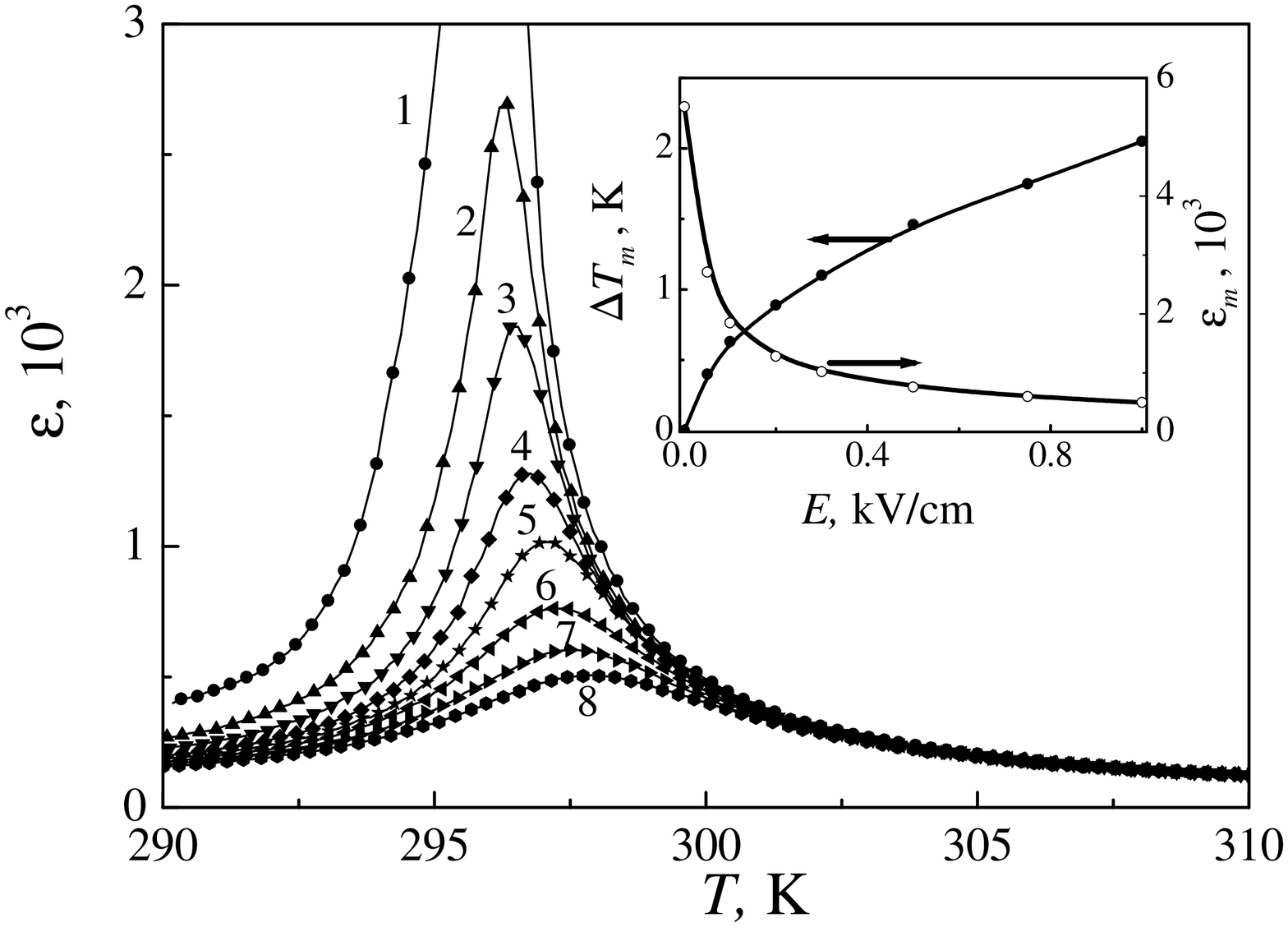}}
\caption{\small Temperature dependences of the dielectric permittivity of Rochelle
salt crystals near upper and lower transition points at different values of
external electric field $E_1$  (kV/cm): 1 --
0, 2 -- 0.05, 3 -- 0.1, 4 --   0.2, 5 -- 0.3, 6 -- 0.5, 7 -- 0.75, 8
-- 1. Lines are guide to the eyes.} \label{slivka5}

\end{figure}

In fig.~\ref{slivka5} we show the measured temperature dependences of dielectric permittivity $\varepsilon_{11}$
of Rochelle salt crystals near the upper and lower transition points at different
values of external d.c. electric field  $E=E_1$ applied along the ferroelectric axis (conjugate to polarization).
The insets contain the field dependences of the dielectric permittivity maxima
$\varepsilon_{m}$ and their temperatures $\Delta
T_{m}=T_{m}(E)-T_{m}(0)$. The data are obtained by cooling samples for the upper maximum and
by heating for the lower one (from the corresponding paraelectric phase towards the ferroelectric phase).
As expected, the external field, conjugate to polarization, decreases the  $\varepsilon_m$
and shifts the maxima temperatures $\Delta T_{m}$ in a non-linear way. For the upper maximum
$\Delta T_{m2} >0$, whereas for the lower one $\Delta T_{m1}<0$.

\begin{figure}[hptb]
\centerline{\includegraphics[width=2.in,angle=270]{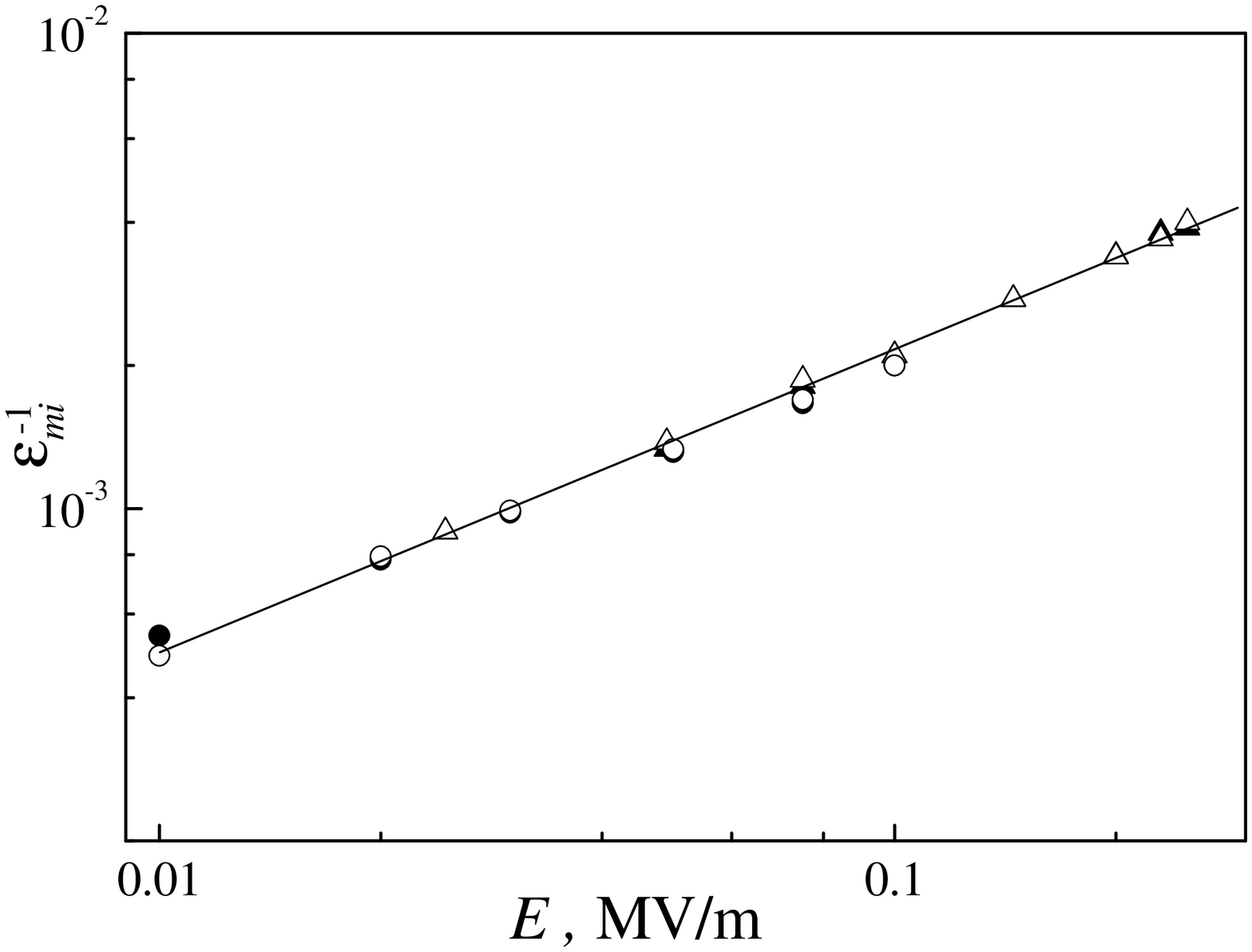}
\hspace{2em}\includegraphics[width=2.in,angle=270]{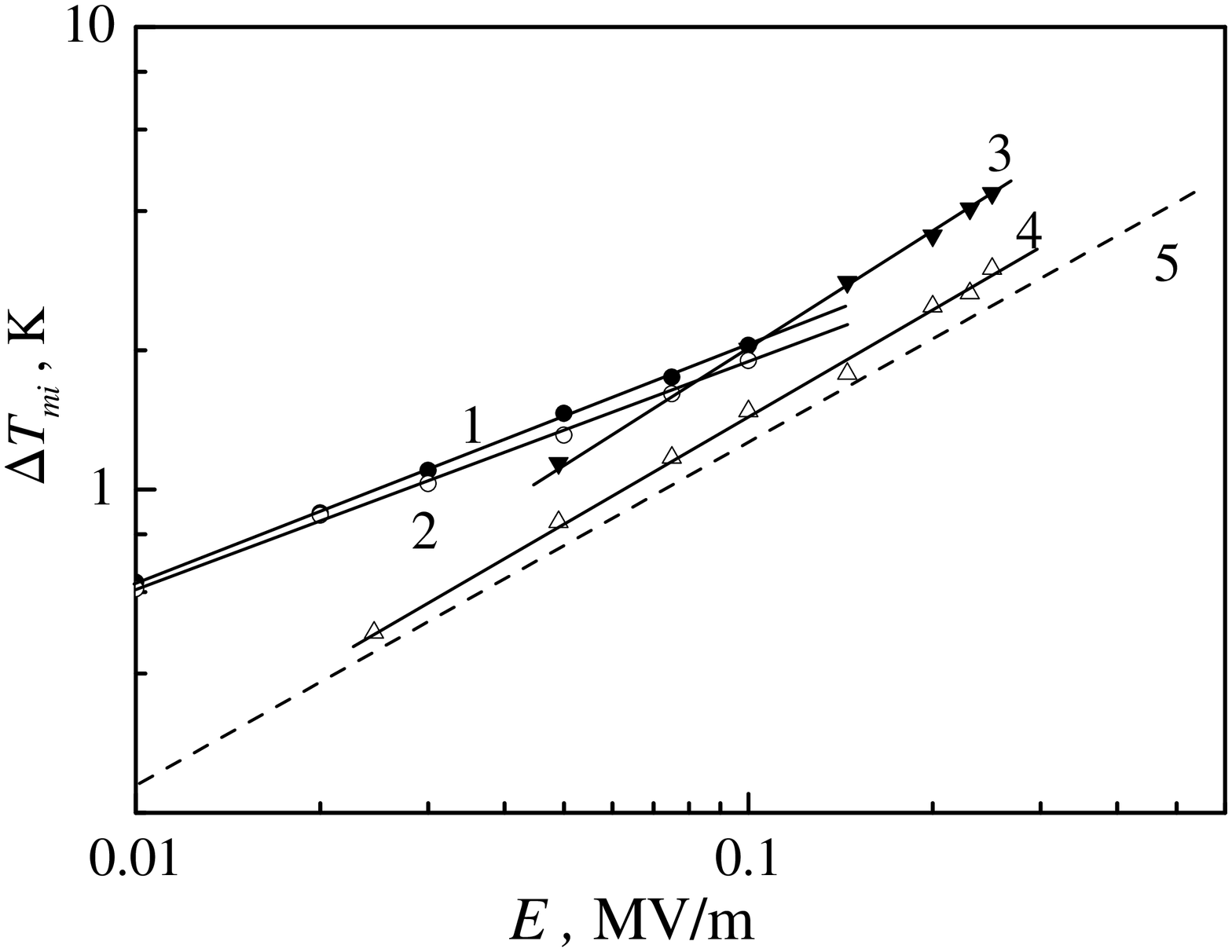}}
\caption{Field dependences of the permittivity maxima magnitudes
(left) and temperature shifts (right). Upper maximum: curve 1 and
$\bullet$ -- this work, curve 3 and $\blacktriangledown$ --
\cite{slivka3}, curve 5 -- \cite{slivka6}. Lower maximum -- curve
2 and $\circ$ -- this work, curve 4 and $\vartriangle$ --
\cite{slivka3}. } \label{slivka6b}
\end{figure}

These results are compared in fig.~\ref{slivka6b} with literature data obtained from the
field dependences of permittivity \cite{slivka2} and elastic compliance $s_{44}^E$
\cite{slivka6}. The obtained in this work field dependences of the
permittivity maxima magnitudes  $\varepsilon_m^{-1}$ are the same for the two maxima (see fig.~\ref{slivka6b})
and well accord with the data of  \cite{slivka2}. However, a perceptible disagreement is observed for
the shift of the permittivity maxima temperatures. Our data
yield very close changes  of $|\Delta T_{m}|$ with field for
the two maxima. On the contrast, the obtained in \cite{slivka2}
field dependence of the upper maximum temperature is much stronger that
of the lower one.

For Rochelle salt the phenomenological Landau expansion of the thermodynamic potential
can be presented as
\begin{equation}
\label{eq1} \Phi(P_1) = \Phi_0 + \frac{\alpha }{2}P_1^2 + \frac{\beta
}{4}P_1^4,
\end{equation}
where $P_1$ is the crystal polarization, $\alpha ,\beta $
are the expansion coefficients. The electric field $E_1$ is applied along the axis of spontaneous polarization [100].

\begin{figure}[htb]
\centerline{\includegraphics[width=2.in,angle=270]{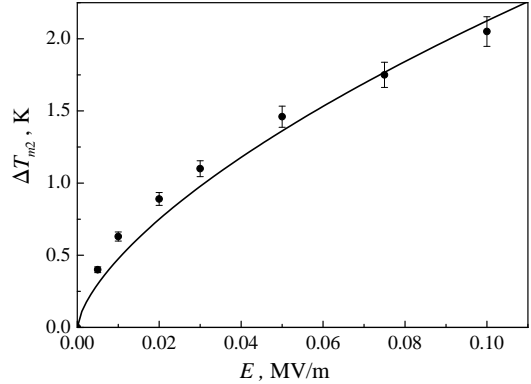}}
\caption{Field dependence of the upper permittivity maximum temperature shift.
Line is calculated with (\ref{eq2}). Symbols are experimental data of this work.} \label{slivka6c}
\end{figure}

For Rochelle salt there are two possible ways to model the temperature dependence of the
coefficient $\alpha$.

1) The expansion (\ref{eq1}) is performed near each of the two transitions separately, assuming
a linear temperature dependence $\alpha = \alpha_{T1} (T_{\rm C1}-T)$ for the lower transition and
$\alpha = \alpha_{T2} (T-T_{\rm C2})$ for the upper one. Then the field dependences of $\varepsilon_m(E_1) $
and $\Delta T_{m}(E_1)$ can be presented as
\begin{equation}
\label{eq3} \varepsilon_m^{ - 1} = \frac32\frac{(4\beta)^{1/3}}{4}\varepsilon_0 E_1^{2 / 3}=k_1E_1^{2 / 3}.
\end{equation}
\begin{equation}
\label{eq2} |\Delta T_{mi}| = \frac34\frac{(4\beta)^{1/3}}{\alpha_{Ti}}E_1^{2 / 3}= k_2E_1^{2 / 3}, \quad i=1,2
\end{equation}

2) Within the second approach, the coefficent
$\alpha $ is chosen in the form
\begin{equation}
\label{eq5} \alpha = \alpha_{1} + \alpha_{2} (T - T_0 )^2,
\end{equation}
where $T_0 = \frac{T_{\rm C1} + T_{\rm C2}}{2}$, and $T_{\rm C1,2} =
T_0 \mp \sqrt{ - \frac{\alpha_{1} }{\alpha_{2} }}$. Such a choice is validated by the fact that
the phase transitions in Rochelle salt are closed to a double critical point \cite{slivka5,slivka6},
realized at partial substitution of potassium atoms with ammonia NH$_{4}$
\cite{slivka9,slivka10}. In  \cite{slivka5,slivka6} the temperature dependences of several
physical characteristics of Rochelle salt were successfully described within the Landau approach with  (\ref{eq5}).

In this case, the field dependences of $\Delta T_{m}(E_1)$ are
\begin{equation}
\label{eq6} \Delta T_{m1,2} = \pm A \mp \sqrt{\frac34\frac{(4\beta)^{1/3}}{\alpha_{T2}}E_1^{2 / 3} + A^2},
\end{equation}
 where $A^2 = - \alpha_{1} /\alpha_{2}$. The field dependence  of $\varepsilon_m(E_1) $
in this case is the same as in the first approach and described by (\ref{eq3}).

The experimental dependences of $\varepsilon_m^{ - 1} (E_1)$ of this work are well described by equation (\ref{eq3})
with $k_1 = 10.2 \cdot 10^{ -7}$(m/V)$^{2 / 3}$ and $\beta =11.34\cdot 10^{13}$ V$\cdot
$m$^{5}$/C$^{3}$. Fitting to the experimental data for $\Delta
T_{m} (E_1)$ with eq.~(\ref{eq2}), shown in Fig.~{slivka6c}, yields the values of $k_2 $ and $\alpha_{T1}$, $\alpha_{T2}$
for the upper and lower maxima:

\[
\noindent {\rm for~~} T_{\rm{C}1}:~~k_{2}=9.9\cdot 10^{ - 4}~{\rm (m/V)^{2
/ 3}~~and~}\alpha_{T1}=5.82\cdot 10^{7}~{\rm V\cdot m\cdot (K\cdot
C)^{ - 1}};
\]
\[
\noindent {\rm for~~} T_{\rm{C}2}:~~k_{2}=10.5\cdot 10^{ -
4}~{\rm (m/V)^{2 / 3}~and~}\alpha_{T2}=5.49\cdot 10^{7}{\rm V\cdot
 m\cdot (K\cdot C)^{ - 1}}.
\]

Agreement with experiment for $\Delta T_{m}(E_1)$, obtained with formulas
(\ref{eq6}) is not any better than with (\ref{eq2}). We found that

$\alpha_{1} = -5.82\cdot10^8$~V$\cdot $m$\cdot $C$^{ - 1}$,~~
$\alpha_{2} = 1.32 \cdot 10^6$~V$\cdot $m$\cdot $C$^{ - 1}\cdot
$К$^{ - 2}$.

Advantages of this approach are visible only at description of the
physical characteristics of Rochelle salt in a sufficiently wide temperature range in the paraelectric phases, where
the non-linear temperature dependence of the inverse permittivity should be essential.  However, for description
of the field dependences of  $\Delta T_{m}(E_1)$ the non-linearity of the coefficient
$\alpha$ within a few Kelvins near the transition points does not play any significant role.

Description of the field dependences of the dielectric permittivity of Rochelle
salt within a modified Mitsui model with piezoeffect
will be given in another publication.

Using the above results, we can estimate the magnitude of internal
bias fields, existing in crystals without annealing and
after 60 min of annealing. In the former and latter cases, the values
of the permittivity at the upper transition point are about
3500 and 5100 (see fig.~\ref{slivka4}). Therefore, using (\ref{eq3}) and the found values of $k_2$, we get
that at the upper Curie point $E_{bias}=0.055$~kV/cm for a non-annealed sample and
$E_{bias}=0.027$~kV/cm for the sample annealed for 60 min.

\section{External pressures}
\subsection{Uniaxial stresses}

\setcounter{equation}{0}
The temperature dependences of dielectric permittivity $\varepsilon_{11}$ of Rochelle
salt were measured at 1 kHz and different values of mechanical
stresses applied along the main crystallographic directions of
unit cell: [100] -- $\sigma_1$, [010] -- $\sigma_2$,
[001] -- $\sigma_3$ and along [011] -- $\tilde\sigma_{4}$. In the reference system with
axes along the main crystallographic directions, the stress $\tilde\sigma_{4}$ can be presented as
\begin{equation}
\label{s4}
\tilde\sigma_{4}=\sigma_{4}+\frac12(\sigma_2+\sigma_3),
\end{equation}
where $\sigma_{4}$ is the shear strain, which for the Rochelle salt symmetry
is the external field conjugate to the order parameter and acts similarly to the electric field $E_1$.

\begin{figure}[htbp]
\centerline{\includegraphics[height=\textwidth,angle=270]{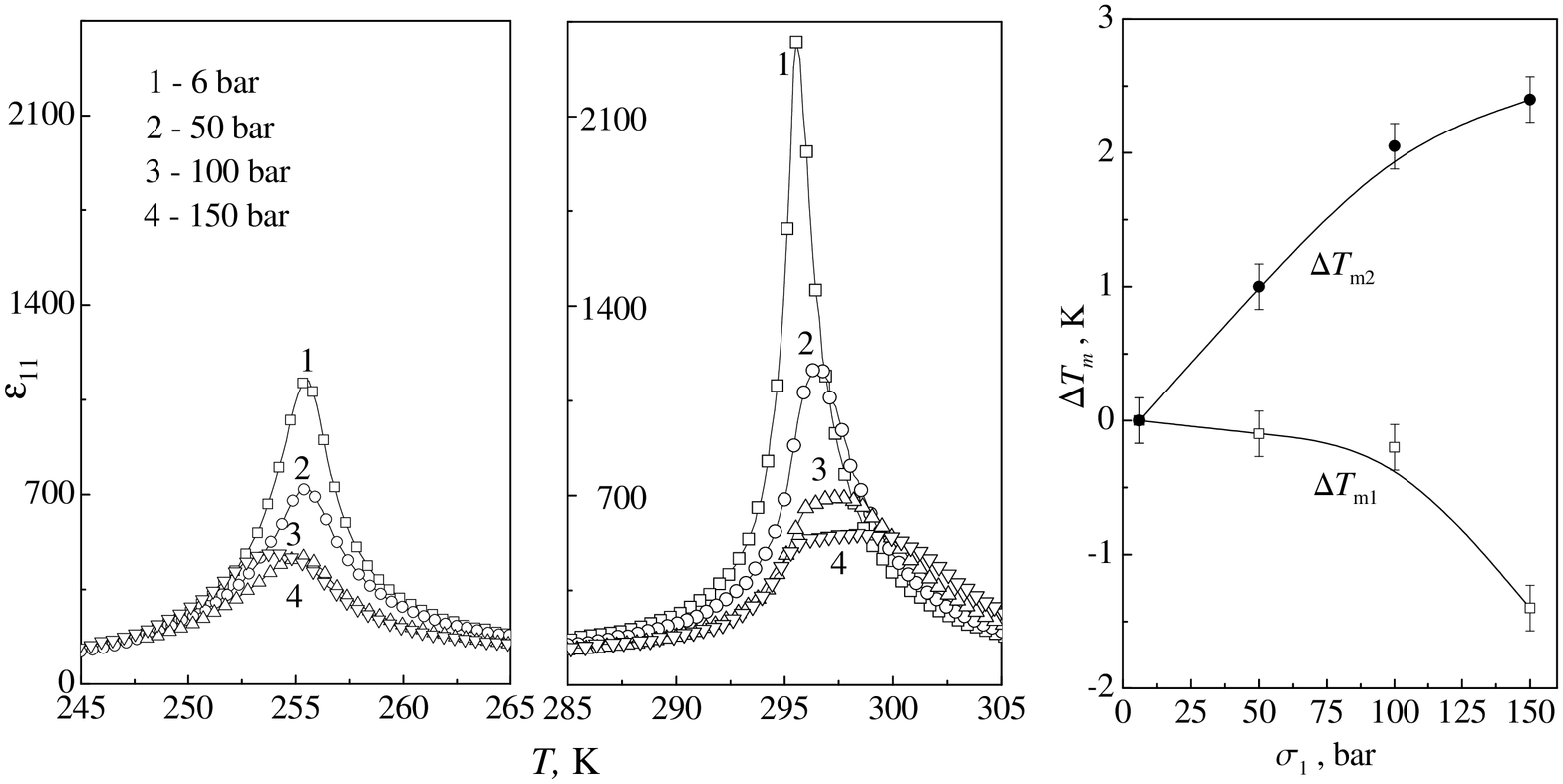}}
\caption{Temperature dependences of the dielectric permittivities of Rochelle salt at different
values of mechanical stress
$\sigma_{1}$ and the stress dependences of the permittivity maxima temperatures.}
\label{slivka7}

\vspace{2em}
\centerline{\includegraphics[height=\textwidth,angle=270]{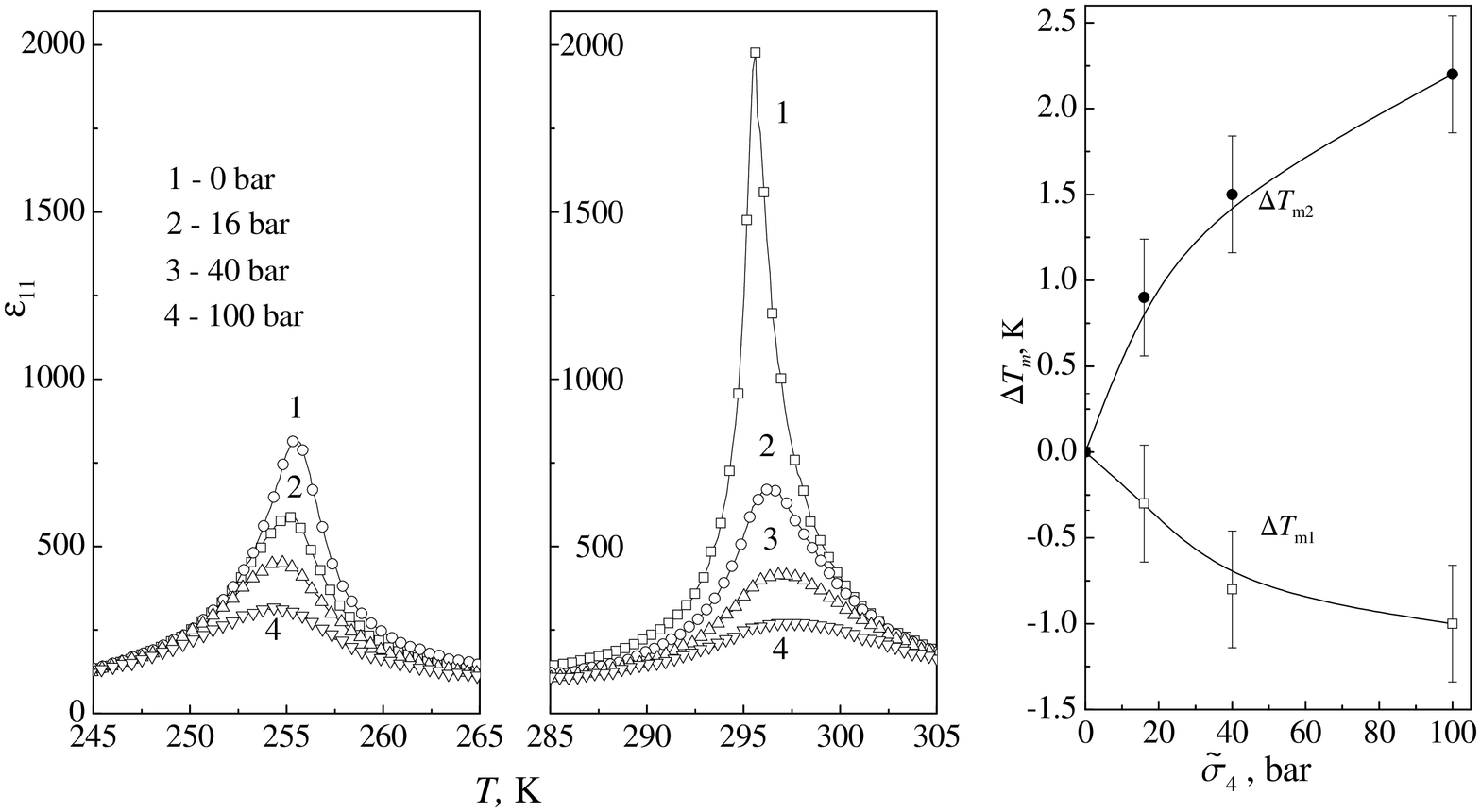}}
\caption{Same for the stress
$\tilde\sigma_{4}$. } \label{slivka8}
\end{figure}

\begin{figure}[htbp]
\centerline{\includegraphics[height=\textwidth,angle=270]{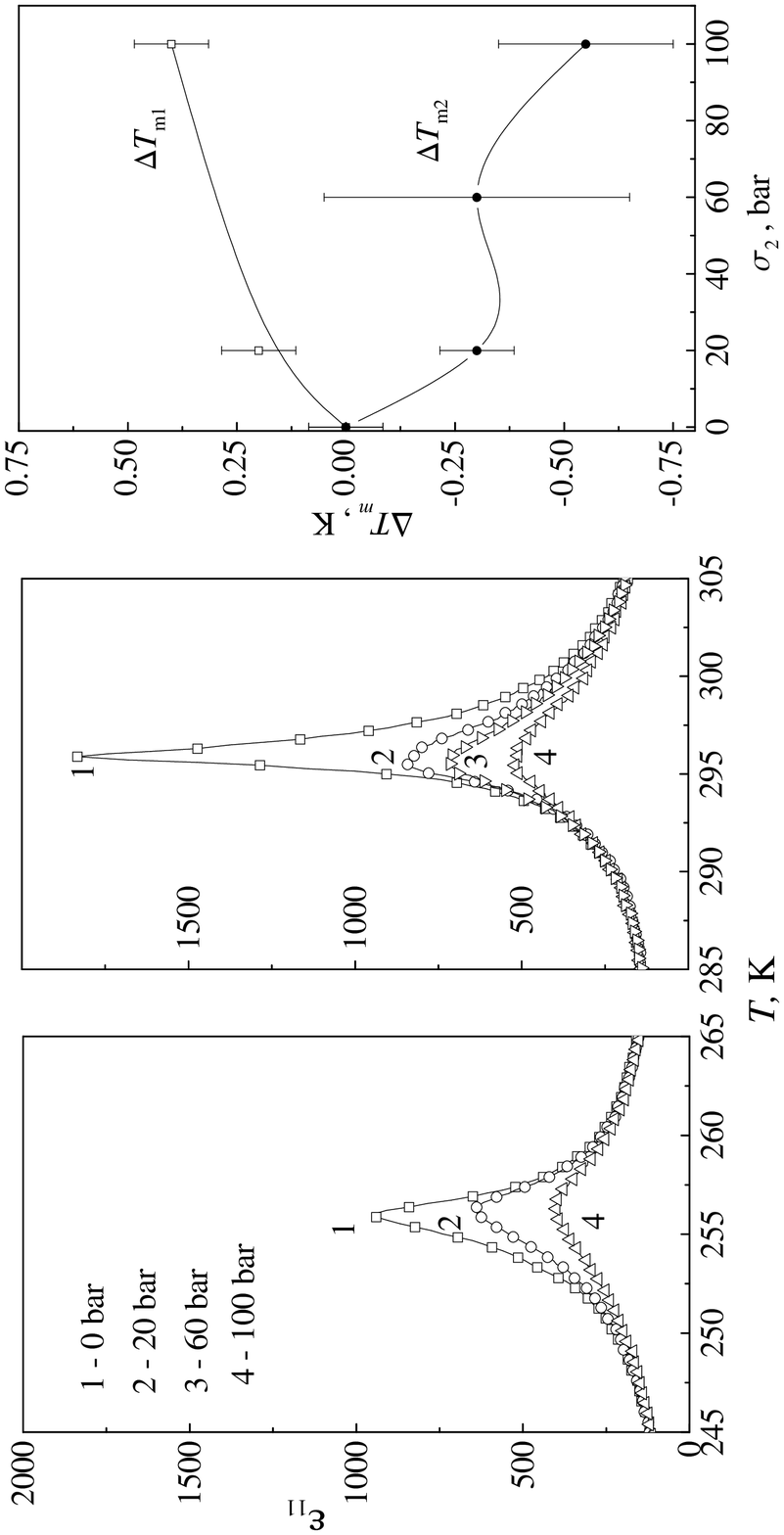}}
\caption{Same for the stress $\sigma_{2}$.}\label{slivka8b}

\centerline{\includegraphics[height=\textwidth,angle=270]{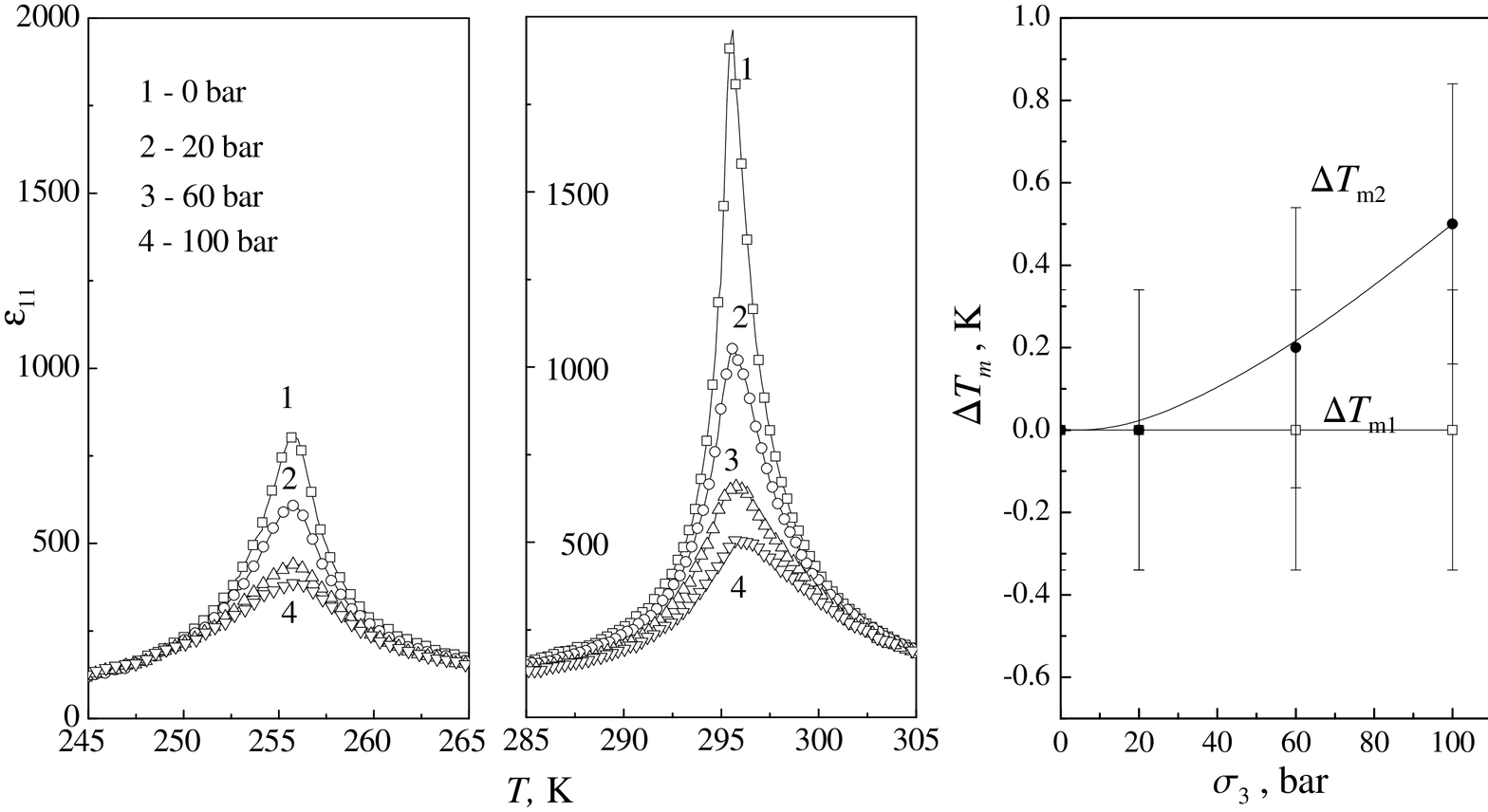}}
\caption{Same for the stress $\sigma_{3}$. } \label{slivka8c}
\end{figure}

Figs.~\ref{slivka7}-\ref{slivka8c} contain the obtained temperature dependences of dielectric permittivites
at different values of the uniaxial pressures and the corresponding stress dependences of the
permittivity maxima temperatures. The data, as in the case of electric field study, were obtained at cooling for
the upper maximum and at heating for the lower maximum (on going from the corresponding paraelectric phase towards
the ferroelectric phase).

Similarly to the electric field $E_1$, all explored uniaxial stresses decrease the maximal
values of the dielectric permittivity and change their temperatures $T_{m1}$ and $T_{m2}$.
Action of the stresses $\sigma_1$ and $\tilde
\sigma_4$ on $T_{m1}$ and $T_{m2}$ is non-linear and similar to the action of the electric field
 $E_1$: $dT_{m1}/d\sigma_i<0$;
$dT_{m2}/d\sigma_i>0$ ($i=1$ and $\tilde4$). It should be noted that the change of the upper maximum temperature
with the stress $\tilde\sigma_{4}$ is much larger than of the
lower maximum. Let us remind that the obtained in this work
changes of the maxima temperatures with the electric field $E_1$ are almost the same for the two
maxima (see figs.~\ref{slivka6b}, \ref{slivka6c}).

%

The experimental data for the shifts of the transition temperature  with
the uniaxial stresses (per 100 bar) are systemized in Table~\ref{table2}.
For comparison, we present here the corresponding literature data
obtained in \cite{imai} on the basis of thermoelastic effect. Overall, the obtained in this work
data qualitatively agree with the literature data, except for the case of stress
$\sigma_3$. However, the quantitative agreement is rather poor,
our data for $|\Delta T_{Ci}|$ being a few times smaller.

\subsection{Hydrostatic pressure}

\begin{figure}[htb]
\centerline{\includegraphics[width=2.4in,angle=270]{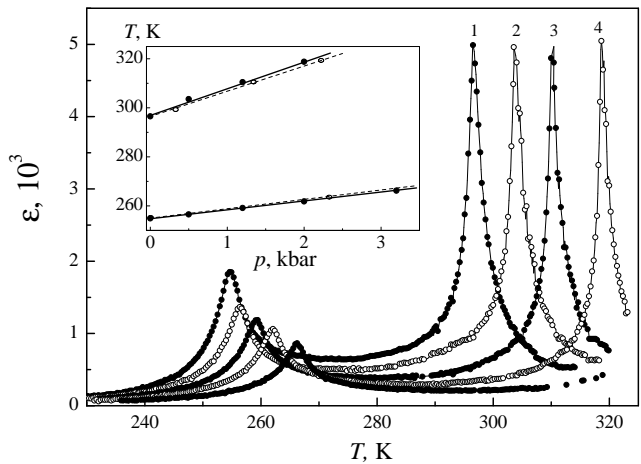}}
\caption{Temperature dependence of the dielectric permittivity of
Rochelle salt at different values of hydrostatic pressure $p$,
MPa: 1 -- 0; 2 -- 50; 3 -- 120; 4 -- 200; 5 -- 320. Inset: the
$p,T$ phase diagram. Dashed lines and $\circ$ -- data of
\protect\cite{slivka8}. } \label{slivka11}
\end{figure}

Fig.~\ref{slivka11} contains the temperature dependences of the dielectric
permittivity of Rochelle salt at different hydrostatic pressures. In contrast
to electric field or uniaxial stresses, the hydrostatic pressure increases both
transition temperatures (see the inset with the  $p,T$-diagram). The pressure coefficients
of the transition temperatures are
$dT_{\rm C1}/dp=3.54$ K/kbar  and $dT_{\rm C2}/dp=10.92$ K/kbar, in a perfect agreement
with the data of \cite{slivka8, slivka11}. On increasing the hydrostatic pressure, the value of
$\varepsilon_{m}$ at the lower transition point $T_{C1}$ monotonously decreases and
remains unchanged at $T_{C2}$.


\subsection{Phenomenological description of pressure effects}
For phenomenological description of external pressure influence on
the phase transitions in Rochelle salt, let us modify the
expansion (\ref{eq1}) in the following way
\begin{equation}
\label{eq1m} \Phi(P_1, \sigma_i) = \Phi_0 + \frac{\alpha }{2}P_1^2 + \frac{\beta
}{4}P_1^4  + \sum_{i=1}^3q_{i1}\sigma_iP_1^2 + g_{14}P_1\sigma_4-\frac12\sum_{ij=1}^4s^P_{ij}\sigma_i
\sigma_j
\end{equation}
(for the sake of simplicity we changed here the signs the stresses $\sigma_i$,
in comparison with the standard notations,
so that the uniaxial compression stresses are positive, and for the hydrostatic
pressure we have $p=\sigma_1=\sigma_2=\sigma_3$. In standard notations $-p=\sigma_1=\sigma_2=\sigma_3$
and  values of the compression stress are negative.)
The quantities $q_{i1}$ have a meaning of electrostriction coefficients,
$s^P_{ij}$ are the elastic compliances at constant polarization. Let us note that
$s^P_{ij}$ for $i=1,2,3$ and $s^P_{44}$ are practically temperature independent, whereas
 $s^P_{i4} \sim P_1$, that is, they are different from zero only
in the ferroelectric phase or in presence of electric field (possibly internal bias field $E_{bias}$ due to
polar defects) or stress $\sigma_4$.

From (\ref{eq1m}) the equations for polarization and lattice strains follow
\begin{eqnarray}
&&E_1=g_{14}\sigma_4+(\alpha+2\sum_{i=1}^3q_{i1}\sigma_i)P_1+\beta
P_1^3-\sum_{i=1}^3\frac{s^P_{i4}}{P_1}\sigma_i\sigma_4\\
&& u_i=\frac{\partial \Phi}{\partial\sigma_i}=-\sum_{j=1}^4s_{ij}^P\sigma_j+q_{i1}P_1^2, \quad i=1,2,3 \\
&& u_4=g_{14}P_1-\sum_{j=1}^4s_{ij}^P\sigma_j.
\end{eqnarray}
Assuming a linear dependence of the coefficient
$\alpha = \alpha_{T1} (T_{\rm C1}-T)$ for the lower transition and
$\alpha = \alpha_{T2} (T-T_{\rm C2})$ for the upper one, we get
for the transition temperatures shift and the inverse values of
the permittivity of a free crystal (at constant stress)
\begin{eqnarray}
&&\Delta
T_{C1,2}=\pm\frac{2}{\alpha_{T1,2}}\sum_{i=1}^3q_{i1}\sigma_i\mp
{k_1}(E_{bias}-g_{14}\sigma_4)^{2/3}\\
&& \label{aa} \varepsilon_{m1,2}^{-1}=k_2(E_{bias}-g_{14}\sigma_4+
\sum_{i=1}^3\frac{s^P_{i4}}{P_1}\sigma_i\sigma_4)^{2/3}.
\end{eqnarray}

Experimental data for $g_{14}$ are rather dispersive (see the systematization in
\cite{slivka1}). We used here the theoretical data for $g_{14}$ of
\cite{slivka1}, agreeing overall with experiment.
The electrostriction coefficients have been determined in \cite{61s5}. We use adjusted here their values,
in order to get a good fit to the hydrostatic pressure dependences
of transition temperatures. The used values of $q_{i1}$, $g_{14}$ at lower and upper
transition points  are given  in table~\ref{table3}.

\begin{table}[hbt]
\caption{The used data for $q_{i1}$ (in m$^4$/C$^2$) and $g_{14}$ (in m$^2$/C).}
\begin{center}
\small
\begin{tabular}{c|ccc|c}
  \hline
&  $q_{11}$ &  $q_{21}$  & $q_{31}$ & $g_{14}$
\\\hline
 $T_{\rm C1}$ &  -7.5 &  4  & 4.5 & $0.174$\\
 $T_{\rm C2}$ &  -10 &  4.3 & 2.5 & $0.195$\\
  \hline
\end{tabular}
\end{center}
\label{table3}
\end{table}

We consider first the case of a perfect crystal ($E_{bias}=0$).
The calculated shifts of the transition temperatures (permittivity
maxima temperatures) with uniaxial and hydrostatic pressures  are
presented in table~\ref{table2}. As one can see, a very good
agreement is obtained with the hydrostatic pressure data, as well
as the data of \cite{imai} for the uniaxial stresses. Agreement
with the calculation data for $\tilde\sigma_4$ is completely
unsatisfactory. Here we used an assumption that, according to
(\ref{s4}), $\tilde \sigma_4=100$~bar corresponds to a sum of
$\sigma_4=50$~bar, $\sigma_2=50$~bar, $\sigma_3=50$~bar. Our data
for the uniaxial stresses are also in a poor agreement with the
phenomenology and with the literature data; however, the data
calculations and of \cite{imai} agree fairly well.

\setlength{\tabcolsep}{4pt}

\begin{table}[hbt]
\caption{Shifts of the transition temperatures with uniaxial stresses (per 100 bar)
and with hydrostatic pressure (per 1 kbar).}
\begin{center}
\small
\begin{tabular}{c|ccc|ccc|ccc|cc|cc}
  \hline
  & \multicolumn{3}{c|}{$\sigma_1$}  & \multicolumn{3}{c|}{$\sigma_2$}  & \multicolumn{3}{c|}{$\sigma_3$} & \multicolumn{2}{c|}{$\tilde \sigma_4$}  & \multicolumn{2}{c}{ hydrostatic} \\
 & exp. & \cite{imai} & calc. &  exp. & \cite{imai} & calc. &
  exp. & \cite{imai} & calc. & exp. &  calc.  & exp.&  calc.\\
  \hline
 $T_{\rm C1}$ & {\bf -1.2} & -2.9 & -2.73  & {\bf 0.4}& 1.5&  1.46  & {\bf 0} &   1.7 & 1.63 & {\bf -1.0} & -8.8   & {\bf 3.43}& 3.64\\
 $T_{\rm C2}$ & {\bf  2.0} & 3.5 & 3.44  & {\bf -0.6} & -1.6 & -1.48  & {\bf 0.5} &-0.8 & -0.86 & {\bf  2.2} & 9.1   & {\bf 10.92} & 10.99 \\
  \hline
\end{tabular}
\end{center}
\label{table2}
\end{table}

It seems likely that the disagreement between the experimental data of this work and of \cite{imai}  should be
attributed to the influence of sample defects. We recalculated the shifts of the maxima temperature with uniaxial pressure with
taking into account also the role of internal bias field, determining them from (\ref{aa}).
A much better agreement was obtained for the stresses $\sigma_2$ and $\sigma_3$:
 $|\Delta T_{Ci}|$ decrease by several times with increasing $E_{bias}=0$. However,
for the stress $\sigma_1$, the presence of the bias field
has further enhanced the theoretical values $|\Delta T_{Ci}|$,
only worsening an agreement with the experimental data of this work.
At the moment, we have no complete explanation of the disagreement
between our experimental data and the data of \cite{imai} and of
the calculations, especially in a view of the fact that for
the hydrostatic pressure a complete coincidence with the literature
data and with phenomenology is obtained.

The reasons for the strong decrease of permittivity maxima magnitude with
 diagonal stresses $\sigma_i$, $i=1,2,3$ is not quite clear either.
As follows from (\ref{aa}), such a decrease can be accounted for by the increase of
the internal bias field $E_{bias}$  or of the coefficient
$k_2 \sim \beta ^{1/3}$. Such an increase of $k_2$ can be obtained if we take into account the
terms of the fourth order of the $\sum_{i=1}^3q_{i1}^{(4)}\sigma_iP_1^4$ type  in the expansion (\ref{eq1m}). Effectively
it would lead to renormalization of the coefficient $\beta\to\beta+4\sum_{i=1}^3q_{i1}^{(4)}\sigma_i$.


%

\section{Conclusions}

\begin{itemize}
\item[-] Strong dependence of the dielectric permittivity of Rochelle salt
on humidity of the storage air is shown. We believe that the
dispersion of experimental data of different literature sources
can be caused by uncontrolled water content during and previously
to the measurements.

\item[-] Dependence of the permittivity value at the transition
points
on duration of thermal annealing in high-temperature paraelectric phase
indicate the existence of internal electric bias fields in the crystals due to point polar defects.

\item[-] Influence of external electric field, uniaxial stresses, and hydrostatic
pressure on the dielectric permittivity is studied. The results are compared with
available literature data. Analysis of the obtained
results is performed within the phenomenological Landau approach.
Possible reasons for discrepancies in the data are discussed.

\end{itemize}

\subsection*{Acknowledgement}

The authors acknowledge support of State Foundation for
Fundamental Studies of Ukraine, project No~02.07/00310.

\end{document}